\documentclass[english,twocolumn]{webofc}
\def \pT {p_\mathrm{T}}
\def \pTcorr {p_\mathrm{T}^{corr}}
\def \pTrec {p_\mathrm{T}^{rec}}
\def \pTemb {p_\mathrm{T}^{emb}}
\def \pTleading {p_\mathrm{T}^{leading}}
\def \kT {k_\mathrm{T}}
\begin{document}
%
\title{Charged jet reconstruction in Au+Au collisions at $\sqrt{s_{NN}}$ = 200 GeV at RHIC}
\def\Title{Charged jet reconstruction in Au+Au collisions at $\sqrt{s_{NN}}$ = 200 GeV at RHIC} 
%
%

\author{Jan Rusnak\inst{1}
}
\def\Author{Jan Rusnak} 

\institute{rusn@email.cz, Nuclear Physics Institute ASCR, Na Truhlarce 39/64, 180 86 Praha 8, Czech Republic}

\abstract{
Jets represent an important tool to explore the properties of the 
hot and dense nuclear matter created in heavy-ion collisions.
However, full jet reconstruction in such events is a challenging 
task due to extremely large and fluctuating background, 
which generates a large population of combinatorial jets 
that overwhelm the true hard jet population.
In order to carry out accurate, data-driven jet measurements over a broad kinematic range in the conditions of small signal
to background ratio, we use several novel approaches in order to measure inclusive charged jet distributions and semi-inclusive charged jet distributions
recoiling from a high $p_{T}$ hadron trigger in central Au+Au collisions at $\sqrt{s_{NN}}$ = 200 GeV.
A very low infrared cutoff on jet constituents of 200 MeV/$c$ is applied in all measurements. 
These jet measurements allow a direct comparison of jet quenching at RHIC and the LHC.
}
\maketitle
%
\section{Motivation} 
\label{intro}
Jets - collimated sprays of hadrons - are well calibrated tools to study the properties of the matter
created in heavy-ion collisions \cite{star_pp}. They are created by fragmentation and hadronization of scattered partons generated in hard momentum exchange in
the initial stages of the collision. While traversing the medium, they interact with the surrounding hot and dense matter
resulting in modification of their fragmentation with respect to the vacuum case (jet quenching)\cite{quenching}. 
This modification of parton fragmentation provides sensitive observables to study properties of
the created matter.

Jet reconstruction in the environment of a high energy nuclear collision is a challenging task, 
due to the large and complex underlying background and its fluctuations within an event which can easily disturb measured jet
distributions. 
In order to overcome the obstacles of jet reconstruction in heavy-ion collisions, we utilize two different methods.
The first method filters out the fake ``combinatorial'' jets by imposing a cut on the transverse momentum of the leading
hadron of each jet. This procedure however imposes a bias on the jet fragmentation.
The second method chooses the hard event by requiring a high momentum hadron trigger. A jet back-to-back to the trigger is then reconstructed.
No cut is imposed on the jet constituents (except a low-$\pT$ cut of 200MeV/$c$) and the jet fragmentation is therefore nearly unbiased.
\section{Analysis}
\label{sec-1}
We have analyzed data from 0-10\% central Au+Au collisons at $\sqrt{s_{NN}}$ = 200 GeV measured by the STAR experiment at RHIC during the run 2011.

Jets are reconstructed using only charged tracks recorded by the STAR Time Projection Chamber (TPC). All tracks are required to have $\pT\geq200$ MeV/$c$.
Implementation of the anti-$\kT$ algorithm in the FASTJET software \cite{fastjet} is used for jet reconstruction. 

The jet resolution parameter $R$ 
is chosen to be $R=0.3$.
The fiducial jet acceptance is then $|\eta|<1-R$ in pseudorapidity and full azimuth.

In the next step, reconstructed jet transverse momentum $\pTrec$ is corrected for the average background energy density

\begin{equation}
 \pTcorr=\pTrec-\rho\cdot A
\end{equation}
 
where $\rho=\mathrm{med}\{{{p^{rec}_{\mathrm{T}, i}}\over{A_{i}}}\}$ is the event-wise median background energy density 
and $A$ is the jet area calculated with the $\kT$ algorithm using the method \cite{jetarea}.

\subsection{Inclusive Jet Analysis}
\label{ijet}

In order to determine the response of the jet to the presence of the highly fluctuating and complex background we embed a simulated 
jet with known transverse momentum $(\pTemb)$ into a real event and calculate $\delta\pT$ as

\begin{equation}
 \delta\pT = \pTrec - \rho\cdot A - \pTemb = \pTcorr - \pTemb
\end{equation}

It was shown, that the $\delta\pT$ distribution is not significantly dependent on the choice of the fragmentation model of the embedded jet \cite{deltapt}. 
With the knowledge of the $\delta\pT$ and with use of a Monte Carlo (MC) generator, a response matrix of the system can be calculated which maps the 
true $\pT$ distribution to the measured one.

A jet momentum distribution is smeared not only by background fluctuations, but also by detector effects. An MC simulation using a parametrization of the TPC tracking efficiency
is used to calculate an approximate detector response matrix.

In order to reduce the combinatorial background, a cut on the transverse momentum of the 
leading hadron  $(\pTleading)$ of the jet is imposed. Also a cut on the jet area \cite{jetarea} $A>0.2$ in case of $R=0.3$ and $A>0.09$ for $R=0.2$ is applied.

In the final step, the measured $\pTcorr$ distribution is corrected for the background and detector effects using an iterative unfolding technique based on Bayes' theorem \cite{bayes}.

\subsection*{Results}
\begin{figure}[h]
\centering
\includegraphics[width=4.5cm]{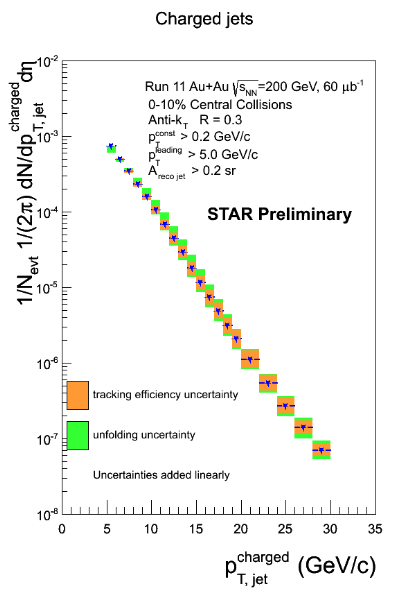}
\caption{The corrected $\pT$ spectrum of inclusive charged jets in central Au+Au collisions at $\sqrt{s_{NN}}$ = 200 GeV for R=0.3.}
\label{fig-incl}       
\end{figure}

Figure \ref{fig-incl} shows the $\pT$ spectrum of inclusive charged jets in central Au+Au collisions at $\sqrt{s_{NN}}$ = 200 GeV for R=0.3 corrected for background and detector effects. 

\subsection{Trigger Recoil Jet Analysis}
\label{ijet}

A trigger hadron is required to have momentum $9\leq\pT\leq19$ GeV/$c$. A jet is then reconstructed in azimuth $\phi$ satisfying
\begin{equation}
 \lvert\phi-\pi\rvert<\frac{\pi}{4}
\end{equation}
where the position $\phi=0$ is defined by the trigger position.

In order to estimate the effect of the presence of the fluctuating background a set of Mixed Events (ME) is created.
A mixed event is composed of $N$ tracks randomly picked up from $N$ different, randomly chosen events (however all the $N$ events come from the same centrality bin, z-vertex bin and event plane direction $\Psi_{EP}$ bin).
All high-$\pT$ tracks are discarded. Such a mixed event does not exhibit any physical correlations between the tracks; 
on the other hand it describes the key features of the background
(detector acceptance inhomogenities, total track multiplicity, etc.).
The mixed event distribution is then subtracted from the (unmixed) Same Event (SE) distribution.

Instead of correcting the results for background and detector effects by means of unfolding, a simulated PYTHIA p+p spectrum is smeared by these effects.
This smeared p+p reference is then compared with the measured Au+Au data.

~
\subsection*{Results}
\begin{figure}[h]
\centering
\includegraphics[width=5.5cm]{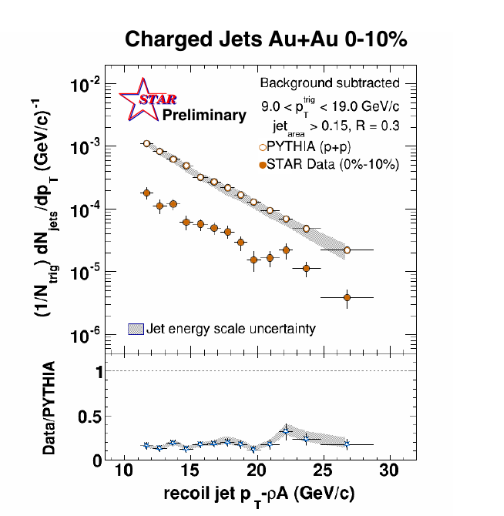}
\caption{The recoil jet spectrum in central Au+Au collisions and smeared PYTHIA p+p spectrum at $\sqrt{s_{NN}}$ = 200 GeV for R=0.3.}
\label{fig-recoil}       
\end{figure}

Figure \ref{fig-recoil} shows a comparison of the measured recoil jet spectrum (SE-ME) in central Au+Au collisions and smeared PYTHIA p+p spectrum for R=0.3.
A suppression of the measured spectrum is apparent with respect to PYTHIA.
%
%
%
\section{Conclusion}
We have presented preliminary results of ongoing jet measurements at the STAR experiment.
These measurements utilize low-bias methods of jet reconstruction allowing direct comparison with theory.

We have used a new technique of the mixed events for jet background estimation in heavy-ion collisions.

Further detector corrections and other effects are yet to be included. 

\section{Acknowledgment}
\small The work has been supported by the MEYS grant CZ.1.07/2.3.00/20.0207 of the European Social Fund (ESF) in the Czech Republic: “Education for Competitiveness Operational Programme” (ECOP).

\end{document}